\documentclass[amsmath,amssymb,twocolumn]{revtex4}
\usepackage{epsfig}
\usepackage{graphicx}
\usepackage{color}

\begin{document}

\title{Sub-Ohmic spin-boson model with off-diagonal coupling: Ground state properties}

\author{Zhiguo L\"{u}$^{1,2}$, Liwei Duan$^1$, Xin Li$^1$, Prathamesh M. Shenai$^1$, Yang Zhao$^1$\footnote{Electronic address:~\url{YZhao@ntu.edu.sg}}}
\date{\today}
\address{$^1$Division of Materials Science, Nanyang Technological University, Singapore 639798, Singapore\\
$^2$Key Laboratory of Artificial Structures and Quantum Control, and Department of Physics, Shanghai Jiao Tong University,
Shanghai 200240, China
}

\begin{abstract}
We have carried out analytical and numerical studies of the spin-boson model in the sub-ohmic regime with the influence of both the diagonal and off-diagonal coupling accounted for via the Davydov ${\rm D}_1$ variational ansatz. While a second-order phase transition is known to be exhibited by this model in the presence of diagonal coupling only, we demonstrate the emergence of a discontinuous first order phase transition upon incorporation of the off-diagonal coupling. A plot of the ground state energy versus magnetization highlights the discontinuous nature of the transition between the isotropic (zero magnetization) state and nematic (finite magnetization) phases. We have also calculated the entanglement entropy and a discontinuity found at a critical coupling strength further supports the discontinuous crossover in the spin-boson model in the presence of off-diagonal coupling. It is further revealed via a canonical transformation approach that for the special case of identical exponents for the spectral densities of the diagonal and the off-diagonal coupling, there exists a continuous crossover from a single localized phase to doubly degenerate localized phase with differing magnetizations.
\end{abstract}

\maketitle

\section{Introduction}

As an archetype of open quantum systems popularized by Leggett \cite{Leggett}, the spin-boson model (SBM) finds applications in condensed matter physics and chemistry in a wide variety of areas ranging from processes of electron transfer \cite{Marcus} to dynamics of qubit-bath entanglement\cite{Khveshchenko,Costi,wang-chen,duanlw}. The SBM consists of a two-level system (TLS) coupled linearly to a bath of harmonic oscillators which accounts for the influence of the environment. The coupling strength between them can be specified by a spectral function $J(\omega)$ which is  proportional to $\omega^s$. Depending on the value of spectral exponent $s$, there exist three distinct cases known as  sub-Ohmic ($s<1$), Ohmic ($s=1$) and super-Ohmic ($s>1$). Theoretical studies predominantly focus on the ground-state properties of the SBM among which the quantum phase transition draws significant attention. No phase transition occurs for the super-Ohmic case, as the system is always in a delocalized ground state. For the Ohmic case in which the SBM can be mapped onto the anisotropic Kondo model by using bosonization techniques, it is well understood that there exists a Kosterlitz-Thouless-type phase transition\cite{Leggett,Weiss}. However, in the most challenging case of the sub-Ohmic regime, a successful and comprehensive description of the SBM has proved elusive despite it being studied by several sophisticated numerical methods\cite{Wilson,Bulla,Makri,Nalbach,Winter,Alvermann,Wang}.

In the sub-Ohmic regime, the slow decay of bath correlations, implying a longer-lasting bath memory effect, renders invalid the methods based on the Born-Markov approximation with even more deteriorating validity in a deeper sub-Ohmic regime with $0<s<0.5$.
The localization-delocalization phase transition for the sub-Ohmic regime has been studied by Bulla \textit{et al.} \cite{Bulla}, wherein non-mean-field exponents for the continuous phase transition in deep sub-Ohmic regime $0<s<0.5$ were obtained. However, recent numerical studies have found mean-field exponents in the deep sub-Ohmic regime as well \cite{Alvermann,Zhang}. By using an extension of the Silbey-Harris variational wave function \cite{sh} which is the two-site version of the Davydov ${\rm D}_1$ ansatz from the theory of ``Davydov soliton'', to study the ground state properties of the sub-Ohmic SBM with $0 < s < 0.5$, Chin {\it et al}.~\cite{Chin} claimed that such a trial wave function generates correct mean-field exponents for the continuous localization-delocalization transition.
It is commonly accepted that there is a second-order phase transition separating a non-degenerate delocalized phase from a doubly degenerate localized phase thanks to a likely bath-induced spontaneous magnetization on the spin \cite{Chin}.  The sub-Ohmic case also shows some special dynamical properties. With an increase in the spin-bath coupling strength, coherent oscillations in the SBM will eventually morph into classical-like damping which is accompanied by the transition from a delocalized state to a localized one \cite{Zheng}. However, recent studies show that the nonequilibrium coherent dynamics can persist for ultrastrong coupling to a sub-Ohmic bath with $s<0.5$ \cite{Kast,WN}.

The SBM \cite{Weiss}
is similar to a one-exciton, two-site version of the Holstein molecular crystal model
\cite{Holstein}, which is among the most popular Hamiltonians for studying optical and transport properties of molecular and biological systems.
In the Holstein Hamiltonian, diagonal coupling is defined as a nontrivial dependence of the exciton site energies on the lattice coordinates, and off-diagonal coupling, as a nontrivial dependence of the exciton transfer integral on the lattice coordinates \cite{SSH}. Off-diagonal interactions between electronic and lattice degrees of freedom have been emphasized by Mahan as modulations of electron-electron
interactions by ion vibrations \cite{mahan}. Similarly, the off-diagonal coupling in the SBM represents bath-induced modulation of the spin tunneling. Accounting for both the diagonal and off-diagonal coupling is important, e.g., in characterizing solid-state excimers, where a variety of experimental and theoretical considerations point to a strong dependence of electronic tunneling upon certain coordinated distortions of neighboring molecules on the formation of bound excited states \cite{non-diagonal1,non-diagonal2}. Mishchenko and Nagaosa have shown that the off-diagonal coupling allows coexistence of free and self-trapped states even in quasi-one-dimensional compound A-PMDA consisting of alternating donor and acceptor molecules \cite{nago2}.
It is well known that there exists no phase transition for the Holstein model if one considers only the diagonal coupling~\cite{Lowen}. However, novel discontinuities have been found in the simultaneous presence of the diagonal and the off-diagonal coupling \cite{Marchand,ZYY}.
It has also been proposed \cite{nago} that the off-diagonal coupling modulates the hopping integral of the Zhang-Rice singlet and the superexchange interaction in the low-doping regime of high-temperature superconductivity. Due to challenges in obtaining reliable solutions \cite{mahan}, Hamiltonians containing off-diagonal coupling appear rarely in the polaron literature, and exciton-phonon interactions in the Holstein model, or the coupling between the system and the bath in the SBM, are usually considered to be of diagonal form.
In this work, by employing the Davydov ${\rm D}_1$ ansatz, we seek to help fill the void in the theoretical study of SBM phase transitions in the deep sub-Ohmic regime $0<s<0.5$ in the simultaneous presence of diagonal and off-diagonal coupling.

The Davydov ansatz and its variants \cite{Scott} have been successfully employed as trial wave functions for the one-dimensional Holstein system \cite{Zhao} and the Rabi model \cite{QH}. The time evolution of a Holstein polaron can be obtained by solving the equations of motion obtained from the Dirac-Frenkel time-dependent variation technique \cite{Dirac,Sun} for the variational parameters of the Davydov ansatz. 
The Davydov ${\rm D}_1$ ansatz has in particular been used as a simple and reliable method to study both the dynamic and static properties of the SBM in deep sub-Ohmic regime $0<s<0.5$ with only diagonal coupling\cite{WN,Chin}.  Furthermore, the finding that the Davydov ${\rm D}_1$ ansatz is especially accurate in the strong exciton-phonon coupling regime \cite{Zhao} from our earlier study, provides support to our contention that the same ansatz may also be reliably employed for SBM in deep sub-Ohmic regime \cite{WN}.

The rest of paper is organized as follows. In Sec.~II, the SBM with simultaneous diagonal and off-diagonal coupling is presented. The Davydov ${\rm D}_1$ ansatz as well as the variational method is also elaborated. In Sec.~III, the numerical results on characterizing the ground state properties are presented and discussed in detail. Conclusions are drawn in the final Sec.~IV.

\section{Methodology}

The SBM Hamiltonian with simultaneous diagonal and off-diagonal coupling can be written as
\begin{eqnarray}
\hat{H}&=&\frac{\varepsilon}{2}\sigma_z-\frac{\Delta}{2}\sigma_x+\sum_l \omega_l b_l^\dag b_l\nonumber\\
\label{Ohami}
&+&\frac{\sigma_z}{2}\sum_l \lambda_l(b^\dag_l+b_l)+\frac{\sigma_x}{2}\sum_l \phi_l(b^\dag_l+b_l).
\end{eqnarray}
where $\hbar=1$ is set to unity, $\sigma_i$ ($i=x, y, z$) are the Pauli matrices of the TLS, $b_l$ ($b_l^{\dagger}$) is the bosonic annihilation (creation) operator, $\varepsilon$ is the bias to describe the influence of the external magnetic field, $\Delta$ is the tunneling amplitude of the TLS, and $\omega_l$ is the phonon frequency while $\lambda_l$ and $\phi_l$ are the corresponding diagonal and off-diagonal coupling strengths respectively. The diagonal and off-diagonal coupling strengths can be determined by the corresponding bath spectral densities $J_z(\omega)$ and $J_x(\omega)$, respectively.
\begin{eqnarray}\label{OspectraZ}
J_z(\omega)=\sum_l \lambda^2_l \delta(\omega-\omega_l)=2\alpha\omega_c^{1-s}\omega^s \Theta(\omega_c-\omega),\\
\label{OspectraX}
J_x(\omega)=\sum_l \phi^2_l \delta(\omega-\omega_l)=2\beta\omega_c^{1-\bar{s}}\omega^{\bar{s}} \Theta(\omega_c-\omega),
\end{eqnarray}
where $\alpha$ and $\beta$ are dimensionless coupling constants, $\Theta(\omega_c-\omega)$ is the Heaviside step function, and $\omega_c$ is the cutoff frequency which is set to be unity throughout this paper. The type of interaction between the TLS and the boson bath is characterized by the spectral exponents $s$ and $\bar{s}$ for diagonal and off-diagonal coupling, respectively. It should be noted that we use a common boson bath for both diagonal and off-diagonal coupling, but with different spectral densities. In the absence of off-diagonal coupling, the term $\varepsilon\sigma_z/2$ in SBM describing the bias between the states $|+\rangle$ and $|-\rangle$ forbids the continuous quantum phase transition.
Introduction of the bias leads to the disruption of the symmetry of the SBM and accordingly, the ground state would tend to be spin-up or spin-down state. This implies the absence of a delocalized state, and that there may exist only a localized state which corresponds to nonzero magnetization.
The off-diagonal coupling denotes the influence of the boson bath on the spin tunneling, and its introduction is believed to compete with the bias. Therefore, it is anticipated that novel features may emerge in the ground state properties.


The Davydov ${\rm D}_1$ ansatz is often employed to describe the motion of an exciton accompanied by a phonon cloud on a finite one-dimensional lattice. Known to be numerically efficient, the hierarchy of Davydov wave functions including the Davydov ${\rm D}_1$ trial state has been widely used to study exciton-phonon dynamics in molecular and biological systems \cite{Zhao,Sun,YJLH2D1}. Recently, Chin \textit{et al.} have demonstrated that the Davydov ${\rm D}_1$ ansatz is capable to describe the ground state of the SBM in the sub-Ohmic regime \cite{Chin}.
While quite accurate results in sub-Ohmic regime ($s\leq0.5$) of the SBM with only diagonal coupling were obtained \cite{Chin}, results based on the same ansatz failed to reproduce the well known phase transition point $\alpha_c=1$ in the Ohmic case~\cite{Chin,CPRB}. We thus focus only on the sub-Ohmic regime when considering both the diagonal and the off-diagonal coupling with $s$ and $\bar{s} \leq 0.5$.
It should further be noted that the original Davydov ${\rm D}_1$ ansatz has been projected onto momentum eigenstates to form Bloch states for the ground-state descriptions of the Holstein polaron \cite{Zhao}.


Aimed at extending the application of the ${\rm D}_1$ ansatz to off-diagonal coupling and studying the ground state properties of SBM, the trial wave function can be given as
\begin{eqnarray}
|{\rm D}_s\rangle&=&A|+\rangle\exp[\sum_l (f_lb_l^\dag-\rm{H.c.})]|0\rangle_{\rm ph}\nonumber\\
\label{trial func}
&+&B|-\rangle\exp[\sum_l (g_lb_l^\dag-\rm{H.c.})]|0\rangle_{\rm ph},
\end{eqnarray}
where $\rm{H.c.}$ stands for Hermitian conjugate, $|+\rangle$ ($|-\rangle$) is the spin up (down) state, and $|0\rangle_{\rm ph}$ is the vacuum state of the boson bath.
$A$ and $B$ are real variational parameters representing occupation amplitudes in states $|+\rangle$ and $|-\rangle$ respectively, and $f_l$ and $g_l$ ($l=1, 2, 3, ...$) label the corresponding  phonon displacements with momentum $\omega_l$. Without loss of generality, $A^2+B^2$, which is the norm of $|{\rm D}_s\rangle$, can be set to unity, and the system energy $E=\langle {\rm D}_s|\hat{H}|{\rm D}_s\rangle$ can then be written as
\begin{eqnarray}
E=&&\frac{\varepsilon}{2}M-\frac{\sqrt{1-M^2}}{2}\bar{\Delta}+\frac{1+M}{2}\sum_l(f_l\lambda_l+f_l^2\omega_l)\nonumber\\
\label{energy}
&-&\frac{1-M}{2}\sum_l(g_l\lambda_l-g_l^2\omega_l),
\end{eqnarray}
where $\bar{\Delta}=\left[\Delta-\sum_l\phi_l(f_l+g_l)\right]\exp\left[-\sum_l(f_l-g_l)^2/2\right]$ is the renormalized tunneling amplitude that involves the modulations of the phonon bath, and $M=\langle {\rm D}_s|\sigma_z|{\rm D}_s\rangle=A^2-B^2$ is the magnetization parameter that will be used in this work to study the critical behavior of SBM.

The ground state of SBM can be obtained by minimizing system energy of Eq.~(\ref{energy}) with respect to the variational parameters $M,f_l$ and $g_l$. Employing the minimization procedure with respect to phonon displacement $f_l$ and $g_l,$ one obtains
\begin{eqnarray}\label{eq f}
(1+M)\left(\omega_lf_l+\frac{\lambda_l}{2}\right)=\frac{\sqrt{1-M^2}}{2}\left(\bar{\phi}_l-\bar{\Delta}(f_l-g_l)\right),\\
\label{eq g}
(1-M)\left(\omega_lg_l-\frac{\lambda_l}{2}\right)=\frac{\sqrt{1-M^2}}{2}\left(\bar{\phi}_l-\bar{\Delta}(g_l-f_l)\right),
\end{eqnarray}
where we have introduced an auxiliary function $\bar{\phi}_l\equiv\phi_l\exp\left[-\sum_l(f_l-g_l)^2/2\right]$.
Combining Eqs.~(\ref{eq f}) and (\ref{eq g}), one arrives at
\begin{eqnarray}
f_l&=&\frac{-\lambda_l\left(M\bar{\Delta}+\sqrt{1-M^2}\omega_l\right)}{2\omega_l\left(\bar{\Delta}+\sqrt{1-M^2}\omega_l\right)}\nonumber\\
\label{eq f1}
&&-\frac{\bar{\phi}_l\left(\sqrt{1-M^2}\bar{\Delta}+(1-M)\omega_l\right)}{2\omega_l\left(\bar{\Delta}+\sqrt{1-M^2}\omega_l\right)},\\
\label{eq g1}
g_l&=&\frac{-\lambda_l\left(M\bar{\Delta}-\sqrt{1-M^2}\omega_l\right)}{2\omega_l\left(\bar{\Delta}+\sqrt{1-M^2}\omega_l\right)}\nonumber\\
&&-\frac{\bar{\phi}_l\left(\sqrt{1-M^2}\bar{\Delta}+(1+M)\omega_l\right)}{2\omega_l\left(\bar{\Delta}+\sqrt{1-M^2}\omega_l\right)}.
\end{eqnarray}
Similarly, energy minimization with respect to the magnetization $M$ yields
\begin{eqnarray}
0=&&\frac{\varepsilon}{2}+\frac{\bar{\Delta}M}{2\sqrt{1-M^2}}+\frac{1}{2}\sum_l(f_l\lambda_l+f_l^2\omega_l)\nonumber\\
\label{eq M}
&+&\frac{1}{2}\sum_l(g_l\lambda_l-g_l^2\omega_l)
\end{eqnarray}

Substituting Eqs.~(\ref{eq f1}) and (\ref{eq g1}) into the system energy of Eq.~(\ref{energy}) accompanied by the spectral densities from Eqs. (\ref{OspectraZ}) and (\ref{OspectraX}), the system energy as a function of the magnetization, ($E(M)$), is obtained. Using the Taylor series expansion for $E(M)$ about $M=0$, we can write:
\begin{equation}
E=c_0+c_1 M+c_2 M^2+c_3 M^3+c_4 M^4+O(M^5),
\label{Taylor_M}
\end{equation}
where $c_i$ are constant coefficients for fixed $\alpha$, $\beta$, $\Delta$ and $\omega_c$. Chin \textit{et al.}~\cite{Chin} found that  in the scaling limit $\omega_c\rightarrow\infty$ for small $M$, the energy expression takes the Landau form without considering the influence of the bias and the off-diagonal coupling. This implies $c_1=0$ and $c_3=0$, and thus, a second-order phase transition exists in the SBM with only the diagonal coupling. Further, the scaling property of the critical coupling $\alpha_c\propto(\Delta/\omega_c)^{1-s}$ is also in good agreement with other numerical approaches for SBM \cite{Bulla,Winter,Alvermann}.

\begin{figure}[tb]
\includegraphics[scale=0.55]{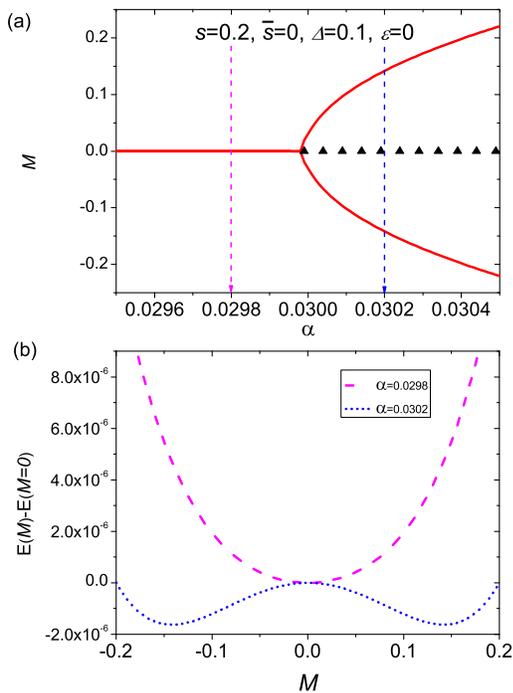}
\vspace{.5cm}
\caption{ For $s=0.2$, $\Delta=0.1$, $\varepsilon = 0$ and without considering the off-diagonal coupling,  (a) the magnetization $M$ corresponding to the extreme values of the system energy as a function of the diagonal coupling strength $\alpha$. The solid red curve corresponds to the minima (ground state) of the system energy, while the curve marked by solid triangles corresponds to energy maxima. (b) The system energy difference $E(M)-E(M=0)$ versus $M$ for two values $\alpha=0.0298$ (dashed, magenta) and $0.0302$ (dotted, blue), which are also depicted by the two vertical arrows in (a) as a guide to eye.}
\label{extreme-diag}
\end{figure}

To the best of our knowledge, the exploration of possible phase transitions in the SBM with the inclusion of off-diagonal coupling has not been systematically undertaken so far, and it thus forms the core of the current work. With the off-diagonal coupling incorporated in the SBM, the expressions of $f_l$ and $g_l$ acquire much more complex forms, presenting difficulties in obtaining analytical results at the phase transition point. The scaling limit in this case yields
\begin{equation}
c_1=\frac{\varepsilon}{2}-\sqrt{\alpha\beta}
\frac{\exp{[-\sum_l(f_l-g_l)^2]}\pi\omega_c}{2\sin(\pi(s+\bar{s})/2)}\left(\frac{\bar{\Delta}}{\omega_c}\right)^{(s+\bar{s})/2},
\label{c1}
\end{equation}
which indicates the absence of a second-order phase transition in general, as $c_1$ and $c_3$ would be non-zero. However, by appropriately selecting $\varepsilon$, $\alpha$, $\beta$ and $\Delta$, it is possible to set $c_1=0$, which may give rise to the possibility of a first-order phase transition as long as $c_3\neq0$. Even though obtaining an analytical expression for $c_3$ is indeed a difficult proposition, it is still possible to judge whether or not $c_3$ is equal to zero using results from numerical analysis as will be elaborated in the next section.

The first requirement in our numerical calculation is to set $c_1=0$. Here we do not consider the Eq.~(\ref{c1}) in our numerical analysis as it is obtained in the scaling limit. An exact expression for $c_1$ can be written as
\begin{equation}
c_1=\frac{dE}{dM}|_{M=0}=\left[\frac{\partial E}{\partial M}+\sum_l(\frac{\partial E}{\partial f_l}\frac{\partial f_l}{\partial M}+\frac{\partial E}{\partial g_l}\frac{\partial g_l}{\partial M})\right]|_{M=0}.
\end{equation}
Therefore, as long as Eqs.~(\ref{eq f}), (\ref{eq g}) and (\ref{eq M}) are satisfied in the condition that $M=0$, $c_1$ would be $0$ and from it we can obtain a relation between the system parameters.
Furthermore, the relaxation iteration technique\cite{Variation2,Variation3} was adopted to obtain numerical solutions to the set of the self-consistency equations [Eqs.~(\ref{eq f}), (\ref{eq g}), (\ref{eq M})] under the condition $c_1=0$. Once the variational parameters are obtained, we can get the ground state wave function, and other properties can be calculated subsequently.

\section{Results and Discussion}

\subsection{Diagonal coupling only}

\begin{figure}[tb]
\includegraphics[scale=0.6]{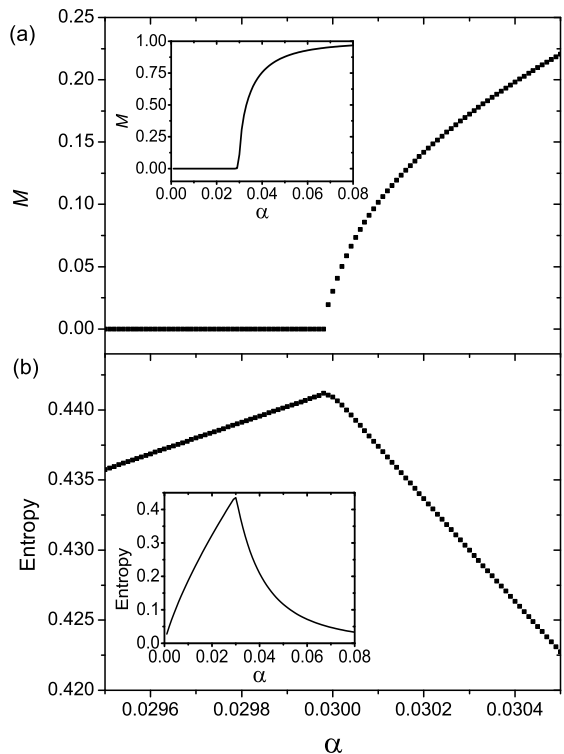}
\vspace{.5cm}
\caption{(a) The positive component of the magnetization ($M$) (the negative part is symmetrically related) and (b) the entanglement entropy as a function of the diagonal coupling strength $\alpha$ for $s=0.2$, $\Delta=0.1$, $\varepsilon = 0$ and without considering the off-diagonal coupling. The inset in each of the panels shows the corresponding plot over a larger range of $\alpha$.}
\label{M-S}
\end{figure}

We first investigate the SBM with only the diagonal coupling considered. In this case, the existence of the second-order transition was demonstrated variationally by Chin \textit{et al.}, and an analytical expression for the phase transition point was also given \cite{Chin}. In this section, we present a detailed description of the magnetization and the ground state energy near the phase transition point for the purely diagonal coupling case by following the approach of Chin \textit{et al.}, for the sake of facilitating an easier distinction with the results upon inclusion of the off-diagonal coupling described in Sec.~III B.

Fig.~\ref{extreme-diag}(a) shows the magnetization corresponding to the extreme values of the system energy as a function of the strength of diagonal coupling $\alpha$ with $s=0.2$ and $\Delta=0.1$ while ignoring the influence of the bias and off-diagonal coupling. There exists a critical coupling strength $\alpha_c=0.02998$ which separates a non-degenerate delocalized phase ($M=0$) from a doubly degenerate localized phase ($M\neq0$) and leads to the bifurcation in the $M(\alpha)$ plot. The solid line in Fig.~\ref{extreme-diag}(a) corresponds to the minima of the ground state energy, while the curve marked by solid triangles for $\alpha > \alpha_c$ corresponds to its maxima. For $\alpha < \alpha_c$, the system energy exhibits only one minimum at $M=0$ and thus characterizes a delocalized state. This is evident in Fig.~\ref{extreme-diag}(b), which shows the detailed energy difference $E(M)-E(M=0)$ as a function of $M$ for a representative value of $\alpha=0.0298$. On the other hand with $\alpha > \alpha_c$, two minima begin to appear in the system energy plot, as can be observed for $\alpha=0.0302$ in Fig.~\ref{extreme-diag}(b). The double-minimum indicates the transition of the system to a localized phase. In this case, $M=0$ still corresponds to an extremum of the system energy, however not being a minimum, it does not correspond to the ground state.



Guided by the magnetization transition, it is straightforward to obtain the phase transition point. As shown in Fig.~\ref{M-S}(a),  the ground state magnetization will change from zero to nonzero at $\alpha_{c} = 0.02998$.
Earlier studies have derived an expression for the dependance of the critical coupling strength on the bath spectral exponent as $\alpha_c\propto(\Delta/\omega_c)^{1-s}$ \cite{Bulla,Winter,Alvermann}.
To further characterize the quantum phase transition in terms of another physical property, we calculate the entanglement between the spin and the surrounding boson bath described by the von Neumann entropy ($S$), also known as the entanglement entropy \cite{Bennett,zz}, which is given for the spin-boson model, as \cite{Costi,Amico}
\begin{equation}\label{entropy}
S=-\omega_+\log_2\omega_+-\omega_-\log_2\omega_-~,
\end{equation}
where
\begin{eqnarray}
\omega_\pm&=&\left(1\pm\sqrt{\langle\sigma_x\rangle^2+\langle\sigma_y\rangle^2+\langle\sigma_z\rangle^2}\right)/2~~.\nonumber\\
\end{eqnarray}
It should be noted that $\langle\sigma_y\rangle=0$ as the spin-boson model is invariant under the transformation $\sigma_y\rightarrow-\sigma_y$~\cite{Costi}.
The calculated entanglement entropy is plotted as a function of $\alpha$ in Fig.~\ref{M-S}(b). With an increase in $\alpha$, $S$ increases gradually until it reaches its maximum at the phase transition point and accordingly the formation of a cusp can be observed clearly in the inset of Fig.~\ref{M-S}(b). We note that in the localized phase, the spin rapidly becomes frozen in one classical state while rapid disentanglement takes place \cite{Costi,Chin}.


\subsection{Simultaneous diagonal and off-diagonal coupling}

The inclusion of the off-diagonal coupling in the Holstein model is known to result in many interesting properties, which leads us to believe that similar implications will also be encountered upon its incorporation in the present work on SBM. Furthermore, due to the similarity between the SBM and the Holstein model, we expect that our current approach of applying the Davydov ${\rm D}_1$ ansatz to the SBM can yield reliable results even upon the inclusion of off-diagonal coupling.

\begin{figure}[tbh]
\includegraphics[scale=0.32]{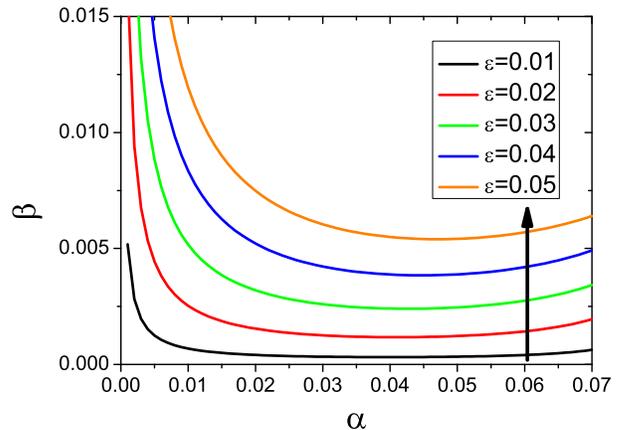}
\caption{The inter-dependence of the diagonal coupling strength $\alpha$ and the off-diagonal coupling strength $\beta$ required in order to satisfy $c_1=0$. Behavior for different values of $\varepsilon=0.01$, $0.02$, $0.03$, $0.04$ and $0.05$ when $s=\bar{s}=0.2$ and $\Delta=0.1$ is shown. The arrow marks the direction of increasing $\varepsilon$. }
\label{a-b}
\end{figure}

\begin{figure}[tbh]
\includegraphics[scale=0.33]{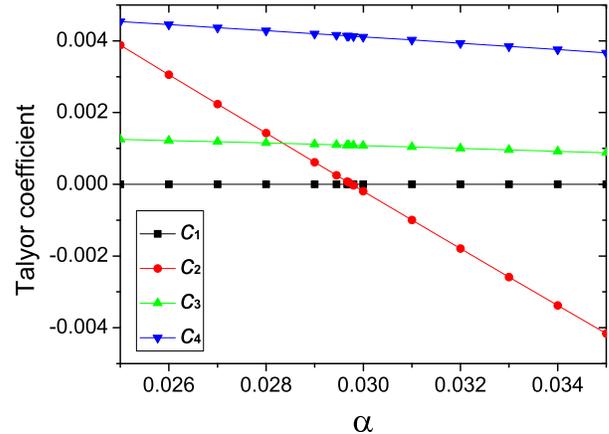}
\caption{The coefficients $c_1$ (squares, black), $c_2$ (circles, red), $c_3$ (up-triangles, green) and $c_4$ (down-triangles, blue) in the Taylor series expansion of the system energy, plotted against the diagonal coupling strength $\alpha$ for $s=\bar{s}=0.2$, $\varepsilon=0.01$ and $\Delta=0.1$.}
\label{derivative}
\end{figure}

As  discussed earlier, in order to set $c_1=0$, a relation has to be established between the diagonal and the off-diagonal coupling strengths. Fig.~\ref{a-b} depicts $\alpha$ as a function of $\beta$ which satisfies the required condition for different values of bias when $s=\bar{s}=0.2$ and $\Delta=0.1$. For a given $\alpha$, the off-diagonal coupling strength $\beta$ can be clearly observed to increase with an increase in the bias intensity. The role of applied bias in localization competes with that of the spin tunneling in delocalization. The modulation of the spin tunneling due to the off-diagonal coupling should thus increase with an increase in the bias. Furthermore, when $\alpha$ tends to zero, $\beta$ shows a tendency to diverge. This behavior emerges since the applied bias will be expected to destroy the possible transition if there is only off-diagonal coupling, as $c_1=0$ cannot be satisfied under such a condition [cf.~ Eq.~(\ref{c1})].

\begin{figure}[tbh]
\includegraphics[scale=0.65]{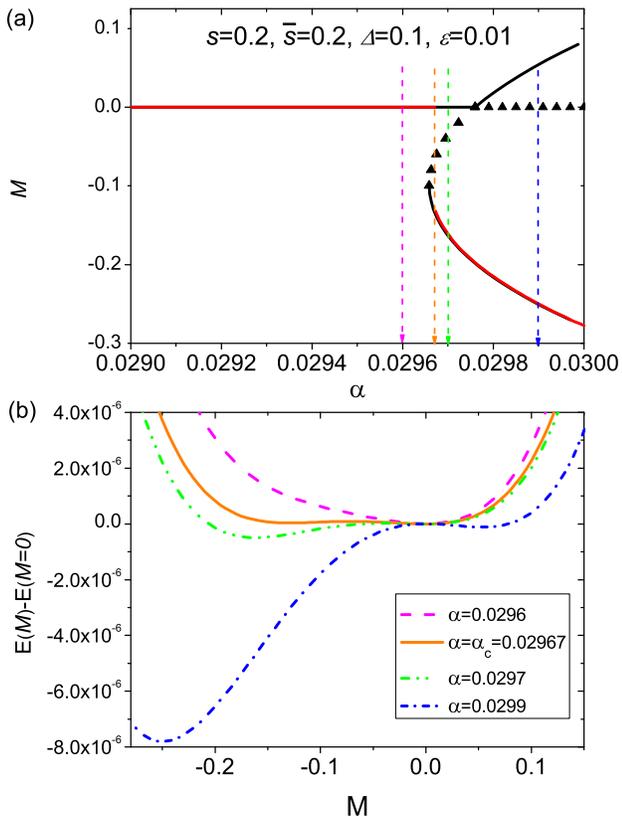}
\vspace{.5cm}
\caption{(a) The magnetization $M$ corresponding to the extreme values of the system energy as a function of the diagonal coupling strength $\alpha$ for $s=\bar{s}=0.2$, $\Delta=0.1$ and $\varepsilon=0.01$. While both the solid curves correspond to local minima of the system energy, only the red curve belongs to the ground state. The curve marked by solid triangles corresponds to the energy maxima.
(b) The system energy difference $E(M)-E(M=0)$ versus $M$ for $\alpha=0.0296$ (dashed, magenta), $0.02967$ (solid, orange), $0.0297$ (dash-double-dotted, green) and $0.0299$ (dash-dotted, blue). As a guide to the eye, the four values of $\alpha$ are marked by vertical arrows in (a).}
\label{extreme-off}
\end{figure}

The coefficients $c_i$ ($i=1, 2, 3, {\rm and}~4$) in the Taylor series expansion as a function of $\alpha$, obtained from the numerical calculations for $s=\bar{s}=0.2$, $\varepsilon=0.01$ and $\Delta=0.1$, are shown in Fig.~\ref{derivative}. The off-diagonal coupling strength $\beta$ changes accordingly to guarantee that $c_1=0$ will always be satisfied. It is important to note here that $c_3$ is generally non-zero, which implies that the second-order phase transition does not exist under these conditions. A first-order phase transition may however still occur, when the condition of $c_2=0$ is satisfied at the critical coupling strength $\alpha_c$.


\begin{figure}[tbh]
\includegraphics[scale=0.65]{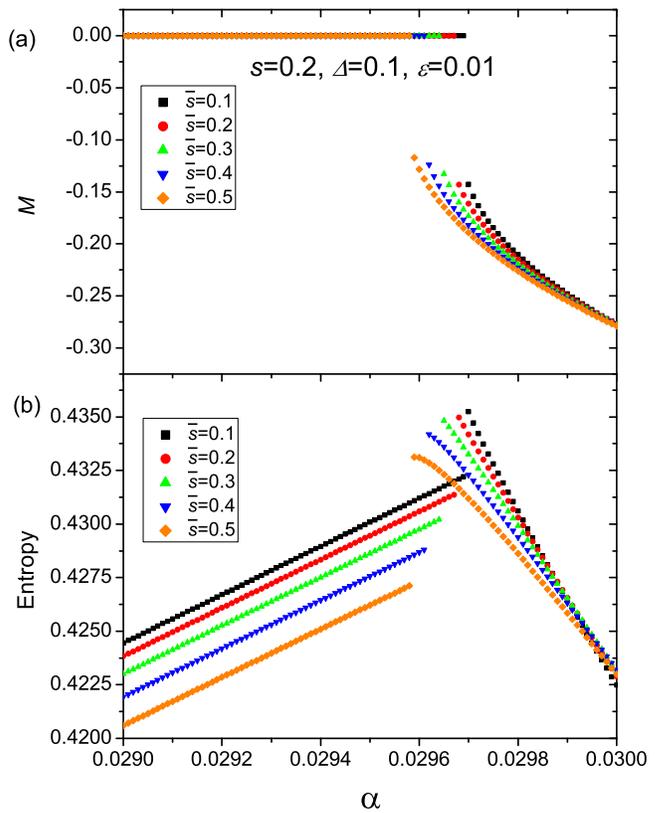}
\vspace{.5cm}
\caption{(a) Magnetization $M$  and (b) entanglement entropy as a function of diagonal coupling strength $\alpha$ for $\bar{s}=0.1$ (squares, black), $0.2$ (circles, red), $0.3$ (up-triangles, green), $0.4$ (down-triangles, blue) and $0.5$ (diamonds, orange) when $s=0.2$, $\Delta=0.1$ and $\varepsilon=0.01$. }
\label{M-S-od}
\end{figure}

The magnetization $M$ corresponding to the extreme value of the system energy as a function of the diagonal coupling strength $\alpha$ for $s=\bar{s}=0.2$, $\varepsilon=0.01$ and $\Delta=0.1$ is shown in Fig.~\ref{extreme-off} (a). While both the solid curves correspond to the minima of the system energy, the red curve represents the ground state of SBM.
The local maxima in the system energy are denoted by solid triangles.
The differences in the magnetization behavior upon inclusion of the off-diagonal coupling as compared to that with only the diagonal coupling can be clearly observed. With the off-diagonal coupling considered, the symmetry about $M=0$ is lost. For $\varepsilon=0.01$ and when $\alpha$ is small, there exists a unique energy minimum at $M=0$. However, when $\alpha$ is large enough, two minima corresponding to different values of $M$ appear. It should be noted that there are two branches, one of which corresponds to $M>0$ showing a continuous change and the other corresponds to $M<0$ showing discontinuous change. Generally, the $M<0$ branch corresponds to the lowest energy and thus to the ground state, which is likely due to the positive bias employed. The ground state of the SBM thus exhibits a discontinuous change in the magnetization from zero to nonzero values, at the critical point, which is found to be $\alpha_c=0.02967$ for the set of parameters chosen here.
In effect, we are witnessing the likely existence of a discontinuous first order transition between a zero-magnetization phase and a finite-magnetization phase of the sub-Ohmic SBM in the presence of off-diagonal coupling. It should be noted, however, that for the special case of $s=\bar{s}$, the SBM Hamiltonian can be subjected to a 90$^{\circ}$ rotation so that the transformed Hamiltonian assumes a form of the diagonal-coupling-only SBM with modified bias and tunneling amplitude. This leads to an ambiguity that
compounds the difficulty of a purely variational approach in proving the existence of a first order phase transition in the SBM.
This warrants a further careful analysis for adopting the notion of a first phase transition at least in the aforementioned special case of $s=\bar{s}$, which is to be presented in the next subsection.
The detailed energy difference $E(M)-E(M=0)$ as a function of $M$ is shown in Fig.~\ref{extreme-off} (b) for four representative values of $\alpha$ including $\alpha_c$. At a less than critical coupling strength $\alpha=0.0296$, a single minimum appears at $M=0$, similar to that of the diagonal case but without any symmetry about $M=0$. With an increase in $\alpha$, another minimum appears at $M<0$ and gradually attains the lowest energy, signifying the ground state.

The ground state magnetization $M$ as a function of the diagonal coupling strength for different $\bar{s}$ is shown in Fig.~\ref{M-S-od} (a) for $s=0.2$, $\Delta=0.1$ and $\varepsilon=0.01$. An abrupt jump in $M$ at critical coupling can be clearly noted. With an increase in $\bar{s}$, the critical diagonal coupling strength decreases slightly. The dependence of the critical point on $\bar{s}$ may be explained by considering the relaxation energy in the SBM given as $\int_0^\infty d\omega{J(\omega)}/{\omega}=2\pi\alpha\omega_c\Gamma(s)$\cite{Lucke}, where $\Gamma(s)$ is the gamma function of $s$ which decreases with an increase of $s$ in the sub-Ohmic regime. If $\bar{s}$ increases, the relaxation energy  will decrease implying that the influence of the off-diagonal coupling will also decrease. Accordingly, we may expect that a smaller value of diagonal coupling strength is needed to balance the off-diagonal coupling.
Furthermore, the entanglement entropy can also be employed to characterize the discontinuous behavior. As shown in Fig.~\ref{M-S-od}(b), the calculated entanglement entropy as a function of $\alpha$ exhibits a discontinuity at the critical coupling. With an $\alpha$ higher than the critical value, the SBM corresponds to the localized state for which rapid disentanglement, similar to that of the purely diagonal case, occurs.
The discontinuous behavior of the magnetization and the entanglement entropy may seem reminiscent of similar discontinuous behavior in the results obtained with the Silbey-Harris method when applied to the biased SBM. Drawing attention to the limitations of the Silbey-Harris ansatz based variational approach, Nazir {\it et al.}~proposed the discontinuous behavior to be regarded as artifacts arising from its excessive simplicity \cite{CPRB}. The Davydov ${\rm D}_1$ ansatz employed in this work, on the other hand, is much more sophisticated and contains more flexible variational parameters as compared to the Silbey-Harris ansatz,  which itself is a special case of the Davydov ${\rm D}_1$ ansatz obtained by setting $A=B$ and $f_l=-g_l$ and is poorly equipped to deal with the asymmetry induced by the bias.




\subsection{Continuous crossover for $s=\bar{s}$}

In this subsection, we carefully analyze the special case when the spectral densities for the diagonal and off-diagonal couplings are characterized by the same exponent. For this condition of $s=\bar{s}$, it is possible to transform the interaction to a purely diagonal form by employing
a unitary matrix $\hat{P} \equiv
\left(
  \begin{array}{cc}
    a & b \\
    b & -a \\
  \end{array}
\right) $
where $a=\sqrt{\frac{1+\lambda_l/g_l}{2}}$, and $b=\sqrt{\frac{1-\lambda_l/g_l}{2}}$. The transformed Hamiltonian can be rewritten as
\begin{eqnarray}
H_{\rm rot}=\frac{\tilde \epsilon}{2}\sigma_z +\frac{\tilde \Delta}{2}\sigma_x+\sum_l\omega_l
b^{\dag}_lb_l+\frac{\sigma_z}{2} \sum_l g_l (b^{\dag}_l+b_l),\label{hami-sb}
\end{eqnarray}
where
\begin{eqnarray*}
  \tilde \epsilon &=& \frac{\sqrt{\alpha} \varepsilon - \sqrt{\beta} \Delta}{\sqrt{\alpha+\beta}}, \\
    \tilde \Delta &=& \frac{\sqrt{\alpha} \Delta + \sqrt{\beta} \varepsilon}{\sqrt{\alpha+\beta}}.
\end{eqnarray*}
The interaction is characterized by the spectral density function $J(\omega)$,
\begin{eqnarray*}
J(\omega) &=& \sum_l g^2_l \delta(\omega-\omega_l)=2(\alpha+\beta)\omega^{1-s}_c\omega^s \Theta(\omega_c-\omega).
\end{eqnarray*}

With this transformation, we have mapped the Hamiltonian (1)
to that of the standard spin-boson model with modified bias and tunneling amplitude via rotation.
Therefore, the ground state of the Hamiltonian (1) is mapped to that of the regular SBM with modified control parameters, i.e., $|G(s,\alpha,\beta,\Delta,\epsilon)\rangle=\hat{P}|G_{\rm{SBM}}(s,\gamma,\tilde{\Delta},\tilde{\epsilon})\rangle$, where $\gamma=\alpha+\beta$.

To obtain the ground state of this model, we adopt an analytical treatment previously developed \cite{lu,sh}, and subject $H_{\rm rot}$ to a unitary transformation in order to take into account the correlation between the spin and bosons, yielding
$H^{\prime }=\exp (S)H_{\rm rot}\exp (-S)$ with
\begin{eqnarray}
 S=\sum_{k}\frac{g_{k}}{2\omega _{k}}(b_{k}^{\dag }-b_{k})[\xi _{k}\sigma_{z}+(1-\xi _{k})\sigma_{0} \textsl{I}].
\end{eqnarray}
Here, $\sigma_0$ is a constant, $\textsl{I}$ is the identity matrix, and $\xi _{k}$ is a $k$-dependent function,
the form of which can be found in Ref.~\cite{lu}. Following the transformation, one can write
\begin{eqnarray}
&&H^{\prime }=H_{0}^{\prime }+H_{1}^{\prime }+H_{2}^{\prime },  \\
&&H_{0}^{\prime }={\frac{\Delta_r}{2}} \sigma_{x}+{\frac{\epsilon^{\prime}}{2}}
 \sigma_{z}+\sum_{k}\omega_{k}b_{k}^{\dag }b_{k}    \nonumber\\
&&-\sum_{k}\frac{g_{k}^{2}}{4\omega_{k}}\xi_{k}(2-\xi
_{k})+\sum_{k} \frac{ g_{k}^{2}}{4\omega_{k}}\sigma _{0}^{2}(1-\xi_{k})^{2}, \nonumber  \\
&&H_{1}^{\prime }={\frac{1}{2}}\sum_{k}g_{k}(1-\xi _{k})(b_{k}^{\dag
}+b_{k})(\sigma_{z}-\sigma_{0}) +i{\frac{\Delta_r}{2}}
\sigma_y B, \nonumber \\
&&H_{2}^{\prime }={\frac{\Delta}{2}}\sigma_x\left( \cosh
\{B\}-\eta \right) +i {\frac{\Delta }{2}}\sigma_y\left( \sinh
\{B\}-\eta B\right), \nonumber
\end{eqnarray}
where $B=\sum_{k} \frac{g_{k}}{\omega_{k}} \xi_{k}(b_{k}^{\dag
}-b_{k})$ and  $\Delta_r= \eta \tilde \Delta$. The renormalized tunneling amplitude is determined as $\eta=\langle\cosh\{B\}\rangle$
(thermodynamically averaged with respect to the Bose-Einstein distribution), thereby yielding
\begin{equation} \label{eta}
\eta =\exp \left[-\sum_{k}\frac{g_{k}^{2}}{2\omega_{k}^{2}}\xi_{k}^{2}\coth\left(\frac{\omega_k}{2T}\right)\right],~ 0
\leq\eta\leq 1.
\end{equation}
Furthermore, the shifted bias is renormalized as
\begin{eqnarray} \label{bias}
&&\epsilon^{\prime}=\tilde{\epsilon}-\tau\sigma_0,
~~~\tau=\sum_{k}\frac{g_{k}^{2}}{ \omega_{k}}(1-\xi_{k})^2.
\end{eqnarray}
It becomes immediately clear that $H_{0}^{\prime}$ can be solved exactly because the spin and the bosons in it are decoupled, and it is easy to obtain the  ground state $|G\rangle $ with energy $E_g$. $H_{0}^{\prime }$ is diagonalized by $U=u\sigma_z+v\sigma_x $ where $u=\sqrt{\left( 1-{\epsilon^{\prime}}/{W}\right)/2}$,
$v=\sqrt{\left( 1+{\epsilon^{\prime}}/{W}\right)/2}$, and
$W=\sqrt{ \epsilon^{\prime 2}+\Delta_r ^{2}}$.
After diagonalization, $\tilde{H}_{0}\equiv U ^{\dag}H'_{0} U $ can be written as
\begin{equation}
\tilde{H}_{0}=\sum_{k} \omega _{k}b_{k}^{\dag }b_{k}
+ \sum_{k} \frac{ g_{k}^{2}}{4\omega_{k}}[ \sigma_{0}^{2}(1-\xi_{k})^{2}-\xi _{k}(2-\xi_{k})]-\frac W 2 \sigma_{z}.\nonumber
\end{equation}
Furthermore, as $\tilde{H}_{1}+\tilde{H}_{2} \equiv U^{\dag }(H_{1}^{\prime }+H_{2}^{\prime }) U $ is treated as a perturbation, the transformation parameters $\sigma _{0}$ and $\xi _{k}$ are chosen in order to minimize $\tilde{H}_{1}+\tilde{H}_{2}$.
Note that in the unitary transformation approach, $\tilde{H}_{1}|G\rangle =0$. Thus, one can write $\tilde{H}_{0}+\tilde{H}_{1}|G_{\rm SBM}\rangle=E_g |G_{\rm SBM}\rangle$, where
\begin{eqnarray} \label{Eg}
E_g=\sum_{k} \frac{g_{k}^{2} }{4\omega _{k}}[(1-\xi_{k})^{2}(1+\sigma_{0}^{2})-1]-\frac{W}{2}.
\end{eqnarray}
The above ground state energy, $E_g$, agrees well with that obtained by the numerical renormalization group
(NRG) method for both zero and finite bias \cite{lu}.
The original Hamiltonian is exactly solvable in two limits, viz., the weak coupling limit of $\alpha\to 0$ and $\beta \to 0$ with $E_g(\alpha\to
0, \beta \to 0)=-{1\over 2}\sqrt{\Delta^2+\epsilon^2}$, and the
zero tunneling limit of $\Delta\to 0$ with $E_g(\Delta\to 0)=-{}|\epsilon|/2-\sum_k g^2_k/4\omega_k$. It is easy to check that $E_g$ yields the correct values of the ground state energy in these two limits.

\begin{figure}
\includegraphics[scale=0.42]{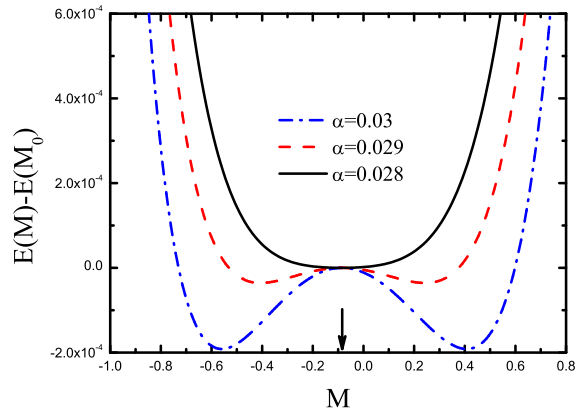}
\caption{ The system energy difference $E(M)-E(M_0)$ versus $M$ for $\alpha=0.028$ (solid, black), $0.029$ (dashed, red), and $0.03$ (dash-dotted, blue) for $s=\bar{s}=0.3$, $\Delta=0.1$ and $\epsilon=0.01 $. }
\label{EgM03D01bs001}
\end{figure}

It is well known that there occurs a continuous phase transition for the unbiased SBM with only the diagonal coupling.
With the introduction of a finite bias ($\varepsilon \neq 0$), and simultaneous diagonal and off-diagonal couplings, the Hamiltonian (1) has no $Z_2$ symmetry {\cite{Braak}}, and thus $\langle \sigma_z \rangle_{G}$ is generally nonzero. However there still exists an instability in the ground state if the modified bias $\tilde\epsilon$ is zero. The calculated energy difference $[E(M)-E(M_0)]$ as a function of $M$($M_0$ corresponds to energy extremum) is shown in Fig.~{\ref{EgM03D01bs001}} for three values of $\alpha$ when $s=\bar{s}=0.3$ and $\Delta=0.1 $. At $\alpha = 0.028$, the system energy exhibits only a single minimum at $M=M_0=-0.086$, and the system is in a localized state. For $\alpha = 0.029$, and $0.03$, it is obvious to see that the system energy exhibits double minima, indicating the instability of the ground state. Furthermore, with an increase in the diagonal coupling strength that is accompanied by a decrease in off-diagonal coupling, there appears a continuous crossover of the SBM from a non-degenerate localized phase to double degenerate localized phase. We note that the necessary condition of the continuous crossover of the ground state is $\tilde{\epsilon}=0$.

In the framework of the transformed SBM with only diagonal coupling, combined coupling $\gamma$ and the modified bias $\tilde \epsilon$ and tunneling amplitude $\tilde \Delta$ all depend on the diagonal and off-diagonal coupling strengths of Hamiltonian (1). As the coupling strengths $\alpha$ and $\beta$ change,  $\tilde \epsilon$, $\tilde \Delta$ and $\gamma$ adopt new values. In the parameter space satisfying the condition $c_1 = 0$ , the modified bias $\tilde \epsilon$ is always nonzero, while it is known that only for the requirement $\tilde \epsilon = 0$, there emerges a continuous crossover of the ground state from single localized phase to a doubly degenerate localized phase. It follows that
the aforementioned continuous crossover does not contradict the discontinuous behavior revealed in the previous subsection using the Davydov ${\rm D}_1$ ansatz, because the condition of $\tilde\epsilon=0$ employed here is inherently different from the requirement of $c_1=0$ (in the expansion for $E(M)$ around $M=0$) in the latter case. From a physical perspective, as the effect of the off-diagonal coupling may run counter to that of the bias, it is possible for the system to stay in a delocalized phase with $M=0$ for a certain off-diagonal coupling strength under a finite bias. As the condition of $c_1=0$ implies that there exists a finite $\tilde \epsilon$, i.e., $\tilde \epsilon \neq 0$, the energy surface $E(M)$ is asymmetric about $M=0$ as shown in Fig.~5. For weak off-diagonal coupling, there is only one minimum at $M=M_0=0$ on the energy surface, and the system is in a delocalized phase.
On the other hand, with $\alpha$ and $\beta$ inter-related as shown in Fig.~\ref{a-b}, for a certain special parameter regime, the energy surface exhibits two asymmetrical localized minima because $\tilde \epsilon \neq 0$, with either a finite $M$ or zero $M$ (cf.~Fig.~5). The minimum with a lower energy corresponds to the stable ground state. At the same time, the maximum occurs at nonzero $M$. This implied the possibility of a phase transition showing an abrupt crossover from a delocalized state with $M=0$ to a localized state with $M \neq 0$ provided that the diagonal and off-diagonal coupling strengths satisfy the condition $c_1=0$. It is thus likely that a discontinuous phase transition occurs as a result of the combined effect of the bias and the two competing forms of the spin-boson coupling, marking a behavior that is entirely different from the continuous phase transition in the unbiased SBM devoid of the off-diagonal coupling. Our work thus reveals the rich energy landscapes and transition properties emerging from the competition between the diagonal and the off-diagonal coupling in the extended sub-Ohmic SBM.

\section{Conclusions and Discussion}

Off-diagonal exciton-phonon coupling is an important issue often neglected by the polaron community.
Early treatments of the off-diagonal coupling in the Holstein Hamiltonian include the Munn-Silbey theory \cite{MunnSilbey,MunnSilbeyZhao}, the Toyozawa ansatz \cite{Zhao},  and Sumi's theory employing the dynamic coherent potential approximation (DCPA) \cite{sumi}. Furthermore, the DCPA based theory was generalized by Kato {\it et al.}~\cite{kato} while an explicit expression was subsequently derived by Hannewald {\it et al.}~for the temperature dependence of the polaron bandwidths by treating diagonal and off-diagonal coupling on an equal footing \cite{VMS}. More recently, the Global-Local Ansatz \cite{GL2}, formulated by Zhao {\it et al.}~in the early 90s, has been compared with DCPA with the Hartree approximation \cite{QM}, and the delocalized form of the Davydov ${\rm D}_1$ ansatz has been used to study the ground-state properties of the Holstein Hamiltonian with off-diagonal coupling \cite{sunjin}.
The absence of phase transitions in the Holstein model when one considers only the diagonal coupling is well known~\cite{Lowen}. Recent studies however indicate that novel nonanalyticities  may emerge in the simultaneous presence of the diagonal and the off-diagonal coupling. For example,
a sharp transition at the critical electron-phonon coupling strength
of the Su-Schrieffer-Heeger model was revealed by Marchand {\it et al.}~\cite{Marchand}.
By employing linearized von Neumann entropy to quantify exciton-phonon correlations in the ground state, Zhang {\it et al.}~uncovered the discontinuities in the presence of off-diagonal coupling of the antisymmetric form as opposed
to the smooth crossover resulting from the symmetric off-diagonal coupling \cite{ZYY}.
Owing to the similarity between the SBM and the Holstein model, trial states from the hierarchy of Davydov ans\"atze may yield reliable results on the ground state properties upon inclusion of the off-diagonal coupling.

In this work we have systematically explored the possibility of a phase transition in the SBM in the simultaneous presence of the diagonal and off-diagonal spin-boson coupling. As the Davydov ${\rm D}_1$ ansatz was employed successfully in studying the Holstein polaron, an analogous approach was taken in this work to investigate the SBM by drawing parallels to its relevance to the SBM. It is demonstrated that a Taylor series expansion of the system energy reveals the possible occurrence of the phase transition as well as its nature, if it does occur. The existence of a continuous phase transition in the SBM with purely diagonal coupling has been shown earlier ~\cite{Chin}, however it is known to vanish in the presence of a bias.
Primarily focussing on the influence of the off-diagonal coupling, the current work reveals the presence of a discontinuous transition between the states characterized by zero and finite magnetization for the sub-Ohmic SBM. It is to be noted, however, that the control parameters $\alpha$, $\beta$, $\Delta$ and $\varepsilon$, must satisfy a certain interrelation to guarantee that that the first order derivative of the system energy with respect to the magnetization $M$ is always zero. This criterion leads to an increase in the required off-diagonal coupling strength with an increase in the bias.
This imposition of restraints on the involved model parameters lends specificity to the resulting trajectory in the parameter space exhibiting discontinuous behavior. Accordingly, the possibility of observing such abrupt transitions in a `realistic' system is rather remote, yet our results bear a significance in yielding valuable theoretical insights on the emergence of novel features in the ground state of the sub-Ohmic SBM.
The magnetization corresponding to the extreme values of the system energy is found to be asymmetrical about zero magnetization. At the critical coupling strength, there exists a discontinuous change from a non-degenerate delocalized state ($M=0$) to another non-degenerate localized state ($M\ne0$). The corresponding critical coupling strength decreases with an increase of $\bar{s}$ if all other control parameters are fixed. We have also probed the entanglement entropy near the phase transition, revealing a discontinuity in it at the transition point.
For $s=\bar{s}$, a unitary transformation has been borrowed to map Hamiltonian (1) to a SBM model with modified bias and tunneling amplitude. In the framework of
the unitary transformation approach \cite{lu}, we have obtained the ground state energy of Hamiltonian (1) and discuss its instabilities. Two types of crossover in the ground state have been uncovered. In addition to the discontinuous crossover from zero magnetization to a finite one, a continuous phase transition also exists in some parameter regime which satisfies zero modified bias ($\tilde \epsilon=0$).
With simultaneously considered diagonal and off-diagonal coupling, this work is thus hoped to shed new light on the emergence of interesting properties in the sub-ohmic regime of the SBM.

We note that in the absence of bias and off-diagonal coupling, the Hamiltonian of Eq.~(\ref{Ohami}) is invariant under the transformation of $\sigma_z\rightarrow -\sigma_z$, $b_l$ $\rightarrow$ $-b_l$ and $b_l^{\dagger}$ $\rightarrow$ $-b_l^{\dagger}$. In this case, there exist two degenerate ground states with differing magnetization characteristics, and their wave functions can then be used to form symmetric and anti-symmetric wave functions which are similar to the Bloch states that are commonly employed in the study of the Holstein model.
Very recently, Bera \textit{et al.} have proposed a trial state~\cite{Bera12} to describe the adiabatic response of the high frequency bosonic modes to the spin tunneling as well as non-classical correlations due to the low frequency modes. With the new ansatz, comparable results to those obtained with the NRG method were obtained, and new behavior in spin coherence and environmental entanglement  was unveiled \cite{Bera12}. Our current approach could well be extended to incorporate the influence of an anti-polaron state via the symmetric and anti-symmetric wave functions. The approach of utilizing such superposed wave functions may be expected to not only allow for improved descriptions of ground state properties in SBM but also lead to revelation of novel properties.



\section*{Acknowledgements}

Support from the Singapore National Research Foundation through the Competitive Research Programme (CRP) under Project No.~NRF-CRP5-2009-04 is gratefully acknowledged. One of us (ZGL) is supported in part by the NSF of China (Grant Nos.~10734020 and
11374208), and the NBRPC (Grant No.~2011CB922202).
The authors thank Lipeng Chen, Qinghu Chen, and Vladimir Chernyak for useful discussion.


\end{document}